\documentclass[aip,apl,preprint,]{revtex4}
\usepackage{amssymb,amsmath}
\usepackage{pgf}
\usepackage{color}
\usepackage{subfigure}
\usepackage{epstopdf}
\usepackage{graphicx}
\usepackage{dcolumn}
\usepackage{bm}
\usepackage[mathlines]{lineno}

\begin{document}

\title[]{Switchable Hyperbolic Metamaterials With Magnetic Control}

\author{Wei Li}
\email{waylee@mail.sim.ac.cn}
\author{Zheng Liu}
\email{liuzheng@mail.sim.ac.cn}
\author{Xiaogang Zhang}
\author{Xunya Jiang}%
 
\affiliation{State Key Laboratory of Functional Materials for Informatics, Shanghai Institute
of Microsystem and Information Technology, Chinese Academy of Sciences, Shanghai 200050,
China}%

\begin{abstract}
A switchable hyperbolic material (SHM) is investigated, with which one can turn on or off
the hyperbolic dispersion of the material via magnetic control. The SHM has simple structure,
with a one-dimensional periodic stacking of dielectric layer and gyromagnetic layer. The hyperbolic
dispersion of SHM is due to the negative effective permeability of gyromagnetic layers, and it can be
transformed into a regular circular dispersion when the d.c. magnetic field is switched off.
This switchable dispersion transition is reversible, which may have great potential applications in many fields.
\end{abstract}

\maketitle

Metamaterials (MMs)\cite{PendryPRL2000,PendrySmithScience2006,LeonhardtScience2006,SoukoulisWegenerScience2010,
ChenShengNM2010} have shown many new striking physics that can be used to control
the electromagnetic properties of materials and go beyond the limit that is attainable
with naturally existing substances. Unlike photonic crystals, MMs can be viewed as
homogeneous media described by effective permittivity $\varepsilon_e$ and effective
magnetic permeability $\mu_e$, since their period is much smaller than the working
wavelength. For anisotropic MMs, their $\varepsilon_e$ and/or $\mu_e$ become tensors.
Recently, an important type of the anisotropic MMs, so called  ``hyperbolic material",
in which one of the diagonal effective permittivity tensor is negative and results in an
anomalous hyperbolic dispersion, has attracted growing attention\cite{JacobOE06,SalandrinoPRB06,
YaoScience08,LiuScience07,ZhangPRL11,YangOL10,LiAPL08,LiuAPL10}. Many interesting phenomena
which are difficult to be realized by natural materials, such as
hyperlensing\cite{JacobOE06,SalandrinoPRB06,YaoScience08,LiuScience07,ZhangPRL11},
all-angle nonreflection\cite{LiAPL08,YangOL10}, and all-direction pulse compression\cite{LiuAPL10}, etc.,
can be achieved by these hyperbolic materials.

The realization of a hyperbolic material has recently advanced 
by adopting the MMs with a periodic metal-dielectric layered structure\cite{SalandrinoPRB06},
or periodic metallic lines\cite{YaoScience08}. For the performance of the hyperbolic material
realized by such structure, one of the most important challenges is the high dissipative losses
of MMs, due to the metallic nature of their constituent meta-molecules. To overcome the losses of MMs,
one can use gain medium for loss-compensation\cite{WuestnerPRL10}. However, this solution need more
complex structures, which may bring new challenges in realization. Besides, this loss-compensation is
frequency-sensitive, which is a drawback for broadband capacity.

On the other hand, the ability to tune and switch the properties of MMs has greatly broaden the applications
of MMs in many fields\cite{ZheludevScience10}, which has been widely studied both experimentally and theoretically
in recent years. Very recently, a ``big flash" event (a large number of photons emit instantaneously,
which has some similarities with the cosmological ``big bang"), is predicated to occur during the metric signature
transition in hyperbolic material\cite{SmolyaninovPRL10}. The metric signature transition in hyperbolic material is
a transition, in which the dispersion relation of the material
is changed between
hyperbolic and elliptic\cite{SmolyaninovPRL10}. Therefore, it is significant to design a switchable hyperbolic material
(SHM), with which one can switch on or off the hyperbolic dispersion.

Reviewing the existing efforts, we think the SHM should include at
least three characteristics: (I) broadband working frequency; (II) low loss; and (III)
switchable hyperbolic dispersion. In this Letter, our design will be presented to serve this purpose.

\begin{figure}
\centering
\includegraphics[width=0.6\columnwidth]{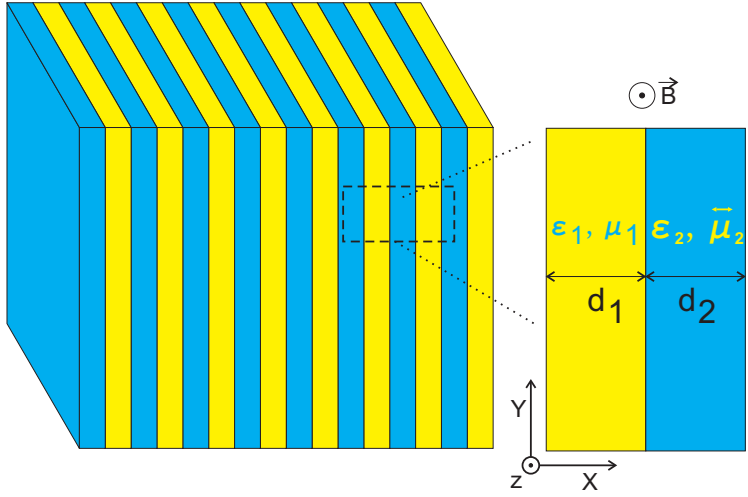}
\caption{\label{} The model of our SHM, which is a one-dimensional periodic stacking of
dielectric layer and gyromagnetic layer, with the thickness $d_1=5.5\mu m$ and $d_2=4.5\mu m$, the relative
permittivity $\varepsilon_1=2.25$ and $\varepsilon_2=15$, the relative permeability $\mu_1=1$,
$\mu_2=\stackrel{\leftrightarrow}{\mu}_2$, respectively. An external uniform d.c. magnetic field is
applied along $z$ direction.}
\end{figure}

The model of our SHM is schematically shown in Fig.1, in which the SHM is a periodic
stacking of dielectric layer and gyromagnetic layer such as yttrium-iron-garnet (YIG) with an external d.c.
magnetic field $H_0$ along $z$ direction. In our model, the thickness of dielectric layer $d_1=5.5\mu m$ and
gyromagnetic layers $d_2=4.5\mu m$ 
are both much smaller than the working wavelengths (about $10^{-2}m$ in this work). Therefore, our
material can be viewed as the effective anisotropic MMs, which can be described by $\varepsilon_e$ tensor
for TM mode or $\mu_e$ tensor for TE mode. In this work, \emph{only} TE mode is considered\cite{ZhangSup10}.
An obvious difference between our structure and the traditional ones\cite{SalandrinoPRB06, LiAPL08} is that we use
the low-loss gyromagnetic material YIG\cite{DasPattonAM2009, lossOfYIG} instead of the high-loss metallic material.That being said, the dissipative losses of our SHM is naturally very low, which can be neglected in this work because both of the dielectric material and the YIG material in the SHM are nearly lossless. Unlike the traditional hyperbolic material, the dispersion of our SHM
is due to the $\mu_e$ tensor, which can be controlled by the external d.c. magnetic field. Therefore,
with magnetic control, one can switch (or tune) the hyperbolic dispersion of the SHM reversibly.

\begin{figure}
\centering
\scalebox{0.7}{\includegraphics{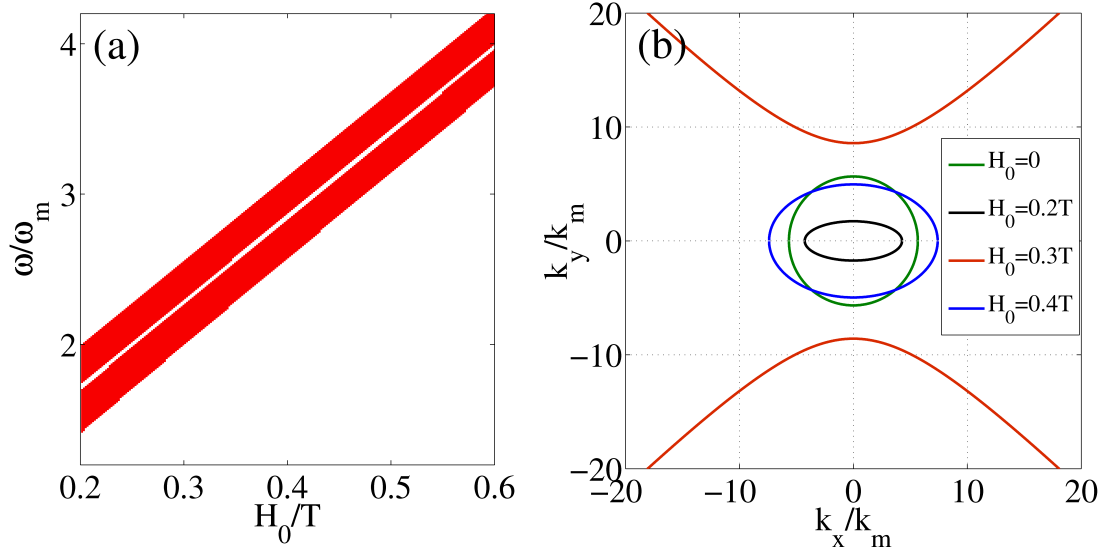}}
\caption{\label{}(a) Existence of hyperbolic dispersion of the SHM at different frequency
$\omega$ with different external d.c. magnetic field $H_0$. The white and the red regions indicate the
elliptic dispersion and the hyperbolic dispersion, respectively. (b) Equifrequency curves in $k$ space at frequency
$\omega=2.02\omega_m$ with different external d.c. magnetic field $H_0$. The green one (circular), the
black one (elliptic), the red one (hyperbolic), and the blue one (elliptic) correspond to $H_0=0$, $H_0=0.2T$,
$H_0=0.3T$, $H_0=0.4T$, respectively.}
\end{figure}

According to real materials, the dielectric constants of dielectric layers and gyromagnetic layers in our
model for the SHM are $\varepsilon_1$=$2.25$, and $\varepsilon_2$=$15$, respectively. With respect to the relative
permeability, the dielectric layers is nonmagnetic with $\mu_1$=1, and the gyromagnetic layers has a
gyromagnetic form\cite{LiuPRB08}:
\begin{equation}\label{permeability}
\stackrel{\leftrightarrow}{\mu}_2=\left[\begin{array}{ccc}
\mu_a&\pm j\mu_b&0\\
\mp j\mu_b&\mu_a&0\\
0&0&1
\end{array}
 \right]
 \end{equation}
where $\mu _a=1+\frac{\omega_m(\omega _0-i\alpha\omega)}{(\omega _0-i\alpha\omega)^2-\omega^2}$,
$\mu _b=\frac{\omega \omega _m}{(\omega _0-i\alpha\omega)^2-\omega ^2}$, $\pm$ and $\mp$ describe
the direction of the external d.c. magnetic field, respectively. $\omega_0=\gamma H_0$ is the resonance
frequency with $\gamma$ as the gyromagnetic ratio. $\alpha$ is the damping
coefficient. The characteristic circular frequency is $\omega_m=5.32GHz$,
corresponding to a wave vector $k_m=\omega_m/c$.
$c$ is the speed of light in vacuum. When $H_0$ is 0.16T, the tensor element in YIG \cite{WangPRL08} is at 4.28 GHz
with $\mu_a=14$ and $\mu_b=12.4$. When $H_0=0$, we have $\mu_a=1$ and $\mu_b=0$.

Using the transfer-matrix method and imposing the Bloch theorem, the dispersion of our SHM associated
with an effective permeability tensor $\mu_e$=$diag(\mu_x,\mu_y,\mu_z)$ can be obtained as
\begin{equation}\label{dispersion}
    \frac{k_x^2}{\mu_{y}}+\frac{k_y^2}{\mu_{x}}=\overline{\varepsilon}\frac{\omega^2}{c^2},
 \end{equation}
where $k_x$ and $k_y$ is the Bloch's wave vector along $x$ axis and the wave vector along $y$ axis,
respectively, with $\overline{\varepsilon}=f_1\varepsilon_1+f_2\varepsilon_2$,
$\mu_{y}=\mu_{z}=f_1\mu_1+f_2\mu_2$, and
\begin{equation}\label{mux}
    \mu_{x}=\left[\frac{f_1}{\mu_1}+\frac{f_2}{\mu_{2e}}-\frac{f_1f_2}{(\mu_1/\mu_{2e})(\mu_b^2/\mu_a^2)}\right]^{-1}.
 \end{equation}
Here $f_1=d_1/(d_1+d_2)=0.55$, $f_2=d_2/(d_1+d_2)=0.45$, and $\mu_{2e}=(\mu_a^2-\mu_b^2)/\mu_a$ is
the effective permeability of the gyromagnetic layers.

From Eq.(\ref{dispersion}), one can find that the dispersion of the SHM becomes hyperbolic when
$\mu_x\mu_y<0$. The sign of $\mu_x\mu_y$ at different frequency with different external d.c. magnetic
field is shown in Fig.2(a), where the white and the red regions indicate positive and
negative $\mu_x\mu_y$, corresponding to the elliptic dispersion and the hyperbolic dispersion, respectively.
From this figure, we can see the hyperbolic dispersion has two frequency ranges (red regions) for each $H_0$.
The width of each range is about $0.28\omega_m$, which is nearly independent on $H_0$. Between two red regions,
there is a very narrow white region with the width of frequency range about $0.03\omega_m$ for each $H_0$,
corresponding to the elliptic dispersion. Therefore, in the SHM, the total width of frequency ranges
that correspond to hyperbolic dispersion is about $0.56\omega_m$ ($\simeq 3.0GHz)$ for each $H_0$,
which is quite a considerable broad range for microwave signal processing.

As an typical example, when $\omega=2.02\omega_m$ and $H_0=0.3T$, we have $\mu_x=2.3$ and $\mu_y=-2.3$,
and the corresponding equifrequency curve is plotted in Fig.2(b) (the red one). Obviously, the dispersion
in this case is hyperbolic.

Why the dispersion of our SHM can exhibit hyperbolic with some $H_0$? Physically, it can be understood
as follows. First, the effective permeability of the gyromagnetic medium $\mu_{2e}$, as a function of $\omega$
and $H_0$, can be negative at some $H_0$ \cite{ZhangSup10,ZhangJiangAPL2012}. Second, with the negative $\mu_{2e}$ of
gyromagnetic layers in our material, its contribution on the hyperbolic dispersion is similar with that of
negative permittivity of metallic layers in the traditional hyperbolic material. As a result, the expression
of $\mu_x$ for the SHM has many similarities with its counterpart for the traditional hyperbolic
materials\cite{SalandrinoPRB06, LiAPL08}, except the last term in brackets of Eq.(\ref{mux}). This term is due to gyromagnetism,
which provides a modification for $\mu_x$ and it disappears in the nongyromagnetic material.

\begin{figure*}
\centering
\includegraphics[width=1.0\columnwidth]{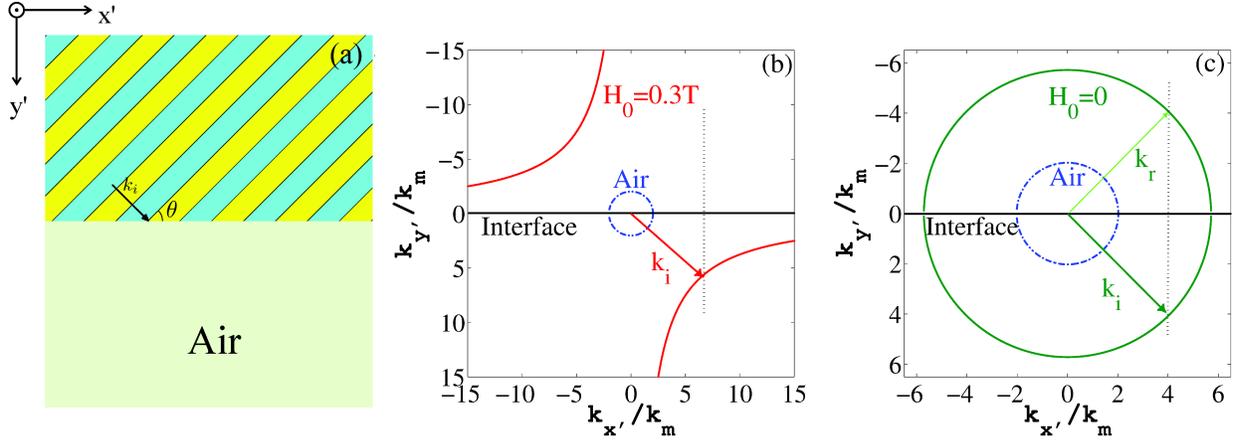}
\caption{\label{} (a)The model for the control of electromagnetic wave on the interface between the SHM and air.
The SHM is terminated with a slanted angle $\theta=\pi/4$, and the beam is incident from the SHM to the air.
(b) and (c) The equifrequency curves and the analysis of reflection and transmission at frequency
$\omega=2.02\omega_m$, with $H_0=0.3T$ and $H_0=0$, respectively. The black solid line indicates the interface, the blue
dashed-dotted one, the red solid one and the green solid one represent the equifrequency curves of air, our SHM with
$H_0=0.3T$ and $H_0=0$, respectively. The arrows indicate the wave vector of the incident wave ($k_i$) and
reflected wave ($k_r$).}
\end{figure*}

Our material is tunable. With magnetic control, the dispersion of our SHM can be transformed,
due to the change of $\mu_e$. As shown in Fig.2(a), for a given frequency, the dispersion of the SHM
is able to be changed between hyperbolicity and ellipticity with the control of $H_0$. To show it more
clearly, the dispersion of the SHM at frequency $\omega=2.02\omega_m$ with $H_0=0$, $H_0=0.2T$, $H_0=0.3T$
and $H_0=0.4T$ are calculated, and the corresponding equifrequency curves are respectively plotted in Fig.2(b).
In this figure, we can find the dispersion of the SHM can be hyperbolic or elliptic, by changing $H_0$.
When $H_0$ is switched off, i.e., $H_0=0$, the dispersion turns into a regular circular one, because our
material becomes nonmagnetic in this case.

Our SHM can be used to realize a reversibly tunable metric signature
transition by switching $H_0$ on and off. This tunable transition may have a great many
potential applications in various fields, i.g., it can be used to control the light on the interface
of the SHM. In our previous work, we have demonstrated that the all-angle zero-reflection \cite{LiAPL08}
and all-direction pulse compression\cite{LiuAPL10} can occur on the interface between the anisotropic medium with
hyperbolic dispersion and the isotropic dielectric. Here we would like to show that these striking phenomena
can be controlled by $H_0$ with our SHM.

For this, we show the control of electromagnetic wave on the interface between the SHM and air.
As shown in Fig.3(a), the surface of the SHM is terminated with a slanted angle $\theta$, and
the electromagnetic wave is incident from the SHM to the air. For the sake of simplicity,
we choose a new Cartesian coordinate system $x'$-$y'$ with the same origin, and the angle between
$x'$ and $x$ axis is ($\pi/2-\theta$). After some algebra, the dispersion of the SHM in the
new coordinate can be obtained as:

\begin{equation}\label{dispersion2}
    \frac{k_{x'}^2\mu_{x'x'}+k_{x'}k_{y'}(\mu_{x'y'}+\mu_{y'x'})+k_{y'}^2\mu_{y'y'}}{\mu_{x'x'}\mu_{y'y'}-\mu_{x'y'}\mu_{y'x'}}=\overline{\varepsilon}\frac{\omega^2}{c^2},
 \end{equation}
where
\begin{equation}\label{}
\begin{split}
\left[\begin{array}{ccc}
\mu_{x'x'},&\mu_{x'y'}\\
\mu_{y'x'},&\mu_{y'y'}
\end{array} \right]=\left[\begin{array}{ccc}
\mu_xs_{\theta}^2+\mu_yc_{\theta}^2,&(\mu_x-\mu_y)s_{\theta}c_{\theta} \\
(\mu_x-\mu_y)s_{\theta}c_{\theta},&\mu_xc_{\theta}^2+\mu_ys_{\theta}^2
\end{array}
 \right]
 \end{split}
 \end{equation}
with $c_{\theta}=\cos\theta$ and $s_{\theta}=\sin\theta$. The dispersion of our SHM at frequency
$\omega=2.02\omega_m$ with $H_0=0$, $H_0=0.3T$ can be obtained from Eq.(\ref{dispersion2}) and the
corresponding equifrequency curve are shown in Fig.3(b) and Fig.3(c), respectively. In this work,
the slanted angle is chosen as $\theta=\pi/4$.

For the slanted angle $\theta$, when a beam from our material with hyperbolic dispersion is incident to the air,
there is a critical condition $\theta=\theta_c=\arctan(|\mu_y/\mu_x|)$ which means the interface ($x'$ axis)
perpendicular to one of the hyperbola-dispersion asymptotes. When $H_0=0.3T$, the critical condition
is $\theta=\theta_c=\pi/4$. At this condition, the beam has zero reflection and zero transmission
for all incident angles, which can be seen from the equifrequency curve analysis of reflection and transmission
on the interface shown in Fig.3(b).  From this figure, we can find that for a incident beam from
the SHM to the air with any angle, the transmitted wave is always evanescent wave,
so the transmitted energy flux is zero,
i.e., zero transmission occurs. Meanwhile, the zero reflection also occurs in this critical condition, for
the reflected beam is absent, as shown in Fig.3(b). With this zero-reflection and zero-transmission effect,
the incident pulses from the SHM with hyperbolic dispersion can be totally compressed or stopped on the interface.

On the other hand, when $H_0$ is switched off, our material become regular circular dispersion, as shown in
Fig.3(c). In this case, the effects of zero reflection, zero transmission, and pulse compression on the
interface are also ``switched off". In Fig.3(c), the ``zero reflection" becomes ``total reflection" with the same
incident beam. Therefore,
one can control the electromagnetic wave on the interface of our material, just by switching $H_0$ on or off.

In reality, here we emphasis that, control of the electromagnetic wave on the interface of our material is
still feasible. Different from the ideal model as discussed above in which the ideal hyperbolic dispersion
is still valid when $k_{x'(y')}\rightarrow\infty$, the strict ``zero reflection" and ``zero transmission" as
well as the completely compressed or stopped pulse on the interface are impossible in reality, since the
hyperbolic dispersion described by Eq.(\ref{dispersion2}) is available only when
$k_{x'(y')}<k_{max}=2\pi/d\simeq 2.8\times 10^{3} k_m$. However, the ``near zero reflection",
``near zero transmission", the strongly compressed and slow pulse\cite{suppl1} could be achieved
easily on the interface of our SHM, because $k_{max}$ in our SHM is a considerable large value
comparing with $k_m$.

To show more about it, a numerical experiment based on the finite-differential-time-domain (FDTD)\cite{eastfdtd} method
is presented, with the calculated model shown in Fig.3(a). An incident Gaussian pulse at the center
frequency $\omega=2.02\omega_m$ from the SHM with $H_0=0.3T$ is just arriving at the interface,
with the $E_z$ field distribution shown in Fig.4(a). When $H_0$ is kept turned on, as shown in Fig.4(b),
the pulse is strongly compressed, and the compressed pulse propagates along the interface very slowly.
Furthermore, the ``near zero reflection" and ``near zero transmission" can also be observed in this dynamical
process.  On the contrast, when $H_0$ is switched off before the pulse arriving at the interface, the pulse is
almost reflected, as shown in Fig.4(c). From this figure, we can see that ``near total reflection" occurs on
the interface, rather than ``near zero reflection", since the dispersion of the SHM in the case of $H_0=0$
is changed into a circular one with the radium larger than that of air, which can also be seen in Fig.3(c).

\begin{figure}
\centering
\includegraphics[width=0.7\columnwidth]{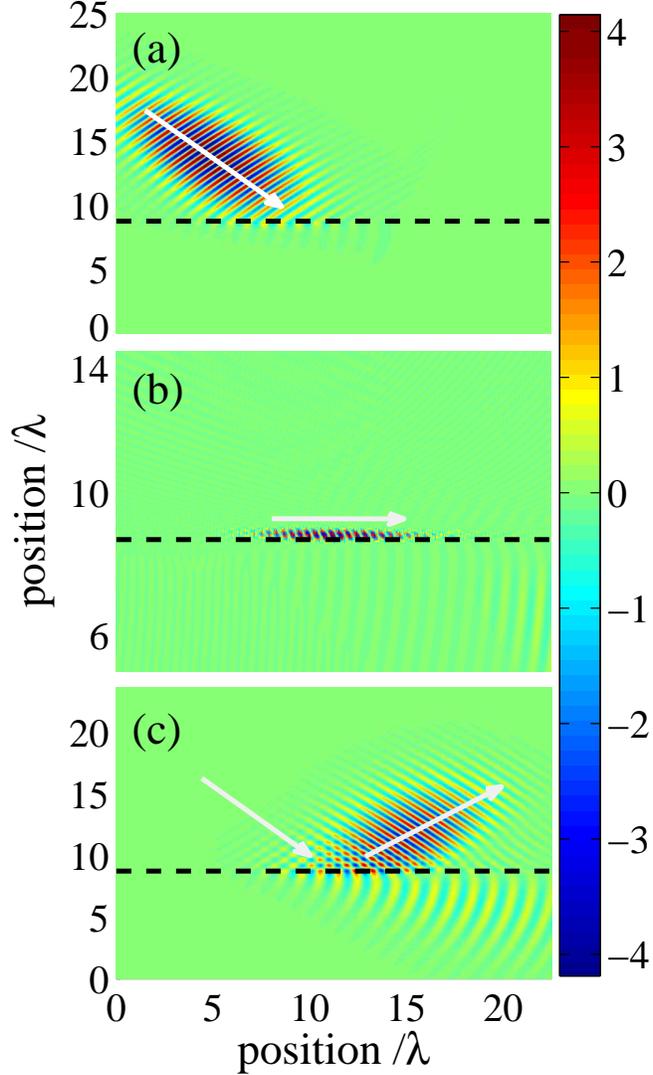}
\caption{\label{}(a) A Gaussian pulse at the center frequency $\omega=2.02\omega_m$ from our material is
incident to the air, with $H_0=0.3T$.  (b) When $H_0$ is kept turned on, the Gaussian pulse is strongly
compressed on the interface, almost without any reflection and transmission. (c) When $H_0$ is switched off
before the Gaussian pulse arriving at the interface, the pulse is nearly totally reflected on the interface.
The white arrows indicate the direction of energy flow of the pulse. The dashed black line indicates the interface.}
\end{figure}

Beyond the example presented in this Letter, there can be much more applications of the SHM,
for instance, it may be used for the observation on ``big flash"\cite{SmolyaninovPRL10}, since the SHM
is very easy to realize the metric signature transition. The related study of ``big flash" in the SHM
will be published in our another paper elsewhere.

In conclusion, in this Letter we have presented our design for the SHM, in which one can
turn on or off the hyperbolic dispersion via magnetic control. The hyperbolic dispersion of the SHM is due to the
negative effective permeability of the gyromagnetic material, and it can be transformed into
a regular circular one when the external d.c. magnetic field is switched off. This tunable dispersion transition is reversible, which may have great potential applications in many fields, such as the control of electromagnetic wave on the interface
of our SHM. In addition, the SHM may be used to observe ``big flash'' since the SHM is very easy to realize the metric signature transition.

\emph{Acknowledgement}. This work was supported by the NSFC (Grant Nos. 11004212, 11174309, and 60938004),
and the STCSM (Grant Nos. 11ZR1443800 and 11JC1414500).


\begin{thebibliography} {s1}
\bibitem{lossOfYIG}The imaginary part of the relative permittivity and the relative permeability of a typical YIG material are both as low as $10^{-3}\sim 10^{-4}$ in microwave region, which are at least $10^{8}$ times smaller than that of a metal such as gold.
\bibitem{suppl1}The group velocity of the compressed pulse on the interface of our SHM with hyperbolic
dispersion is very slow, with the slow limit $v_{gx'}$ (perpendicular to the interface) and $v_{gy'}$ (parallel to the interface) satisfying:
$v_{gx'}=\partial \omega/\partial k_{x'} \propto 1/\gamma_s$ and $v_{gy'}=\partial \omega/\partial k_{y'}\propto 1/\gamma_s^2$, where $\gamma_s=k_{max}/k_0=\frac{1}{d_1+d_2}\frac{2\pi}{k_0}$
is the slowing coefficient. See also Ref.\cite{LiuAPL10}.
\end{thebibliography}

\end{document}